\documentclass[iop]{emulateapj}

\def\C10{C10b}
\def\be{\begin{equation}} 
\def\ee{\end{equation}}

\def\gsim{\lower.5ex\hbox{\gtsima}} 
\def\lsim{\lower.5ex\hbox{\ltsima}} \def\gtsima{$\; \buildrel > \over 
\sim \;$} \def\ltsima{$\; \buildrel < \over \sim \;$} \def\prosima{$\; 
\buildrel \propto \over \sim \;$} \def\gsim{\lower.5ex\hbox{\gtsima}} 
\def\lsim{\lower.5ex\hbox{\ltsima}} 
\def\simgt{\lower.5ex\hbox{\gtsima}} 
\def\simlt{\lower.5ex\hbox{\ltsima}} 
\def\simpr{\lower.5ex\hbox{\prosima}}   
  
 \def\gtsima{$\; \buildrel > \over \sim \;$} 
\def\ltsima{$\; \buildrel < \over \sim \;$} 
\def\gsim{\lower.5ex\hbox{\gtsima}} 
\def\lsim{\lower.5ex\hbox{\ltsima}} 
\def\simgt{\lower.5ex\hbox{\gtsima}} 
\def\simlt{\lower.5ex\hbox{\ltsima}} 
\def\simpr{\lower.5ex\hbox{\prosima}}

\def\E3{{\cal E}_{\rm g}^{III}}

\shorttitle{Optical line emission at z$\sim$6.8}
\shortauthors{M. Castellano et al.}

\begin{document}


\title{Optical line emission from z$\sim$6.8 sources with deep constraints on Ly$\alpha$ visibility}

\author{M. Castellano\altaffilmark{1}, L. Pentericci\altaffilmark{1}, A. Fontana\altaffilmark{1}, E. Vanzella\altaffilmark{2}, E. Merlin\altaffilmark{1}, S. De Barros\altaffilmark{2,3}, R. Amorin\altaffilmark{1},  K. I. Caputi\altaffilmark{4}, S. Cristiani\altaffilmark{5},  S. L. Finkelstein\altaffilmark{6}, E. Giallongo\altaffilmark{1}, A. Grazian\altaffilmark{1}, A. Koekemoer\altaffilmark{7}, R. Maiolino\altaffilmark{8,9},  D. Paris\altaffilmark{1},  S. Pilo\altaffilmark{1}, P. Santini\altaffilmark{1}, H. Yan\altaffilmark{10}}

\altaffiltext{1}{INAF - Osservatorio Astronomico di Roma, Via Frascati 33, I - 00040 Monte Porzio Catone (RM), Italy}
\altaffiltext{2}{INAF - Osservatorio Astronomico di Bologna, Via Ranzani 1, I - 40127, Bologna, Italy}
\altaffiltext{3}{Observatoire de Gen\`{e}ve, Universit\'{e} de Gen\`{e}ve, 51 Ch. des Maillettes, 1290, Versoix, Switzerland}
\altaffiltext{4}{Kapteyn Astronomical Institute, University of Groningen, Postbus 800, 9700 AV Groningen, The Netherlands}
\altaffiltext{5}{INAF Osservatorio Astronomico di Trieste, via G. B. Tiepolo 11, I-34143, Trieste, Italy}
\altaffiltext{6}{Department of Astronomy, The University of Texas at Austin, C1400, Austin, TX 78712, USA}
\altaffiltext{7}{Space Telescope Science Institute, 3700 San Martin Drive, Baltimore, MD 21218, USA}
\altaffiltext{8}{Cavendish Laboratory, University of Cambridge, 19 J. J. Thomson Ave, Cambridge CB3 0HE, UK }
\altaffiltext{9}{Kavli Institute for Cosmology, University of Cambridge, Madingley Road, Cambridge CB3 0HA, UK}
\altaffiltext{10}{Department of Physics and Astronomy, University of Missouri-Columbia, Columbia, MO, USA}

\email{marco.castellano\char64oa-roma.inaf.it}

\begin{abstract}
We analyze a sample of $z$-dropout galaxies in the CANDELS GOODS South and UDS fields that have been targeted by a dedicated spectroscopic campaign aimed at detecting their Ly$\alpha$ line. Deep IRAC observations at 3.6 and 4.5 $\mu$m are used to determine the strength of optical emission lines affecting these bands at z$\sim$6.5-6.9 in order to i) investigate possible physical differences between Ly$\alpha$ emitting and non-emitting sources; ii) constrain the escape fraction of ionizing photons; iii) provide an estimate of the specific star-formation rate at high redshifts. We find evidence of strong [OIII]+H$\beta$ emission in the average (stacked) SEDs of galaxies both with and without Ly$\alpha$ emission. The blue IRAC [3.6]-[4.5] color of the stack with detected Ly$\alpha$ line can be converted into a rest-frame equivalent width EW([OIII]+H$\beta$)=1500$^{+530}_{-440}$\AA~assuming a flat intrinsic stellar continuum. This strong optical line emission enables a first estimate of f$_{esc}\lesssim$20\% on the escape fraction of ionizing photons from Ly$\alpha$ detected objects.  The objects with no Ly$\alpha$ line show less extreme EW([OIII]+H$\beta$)=520$^{+170}_{-150}$\AA~suggesting different physical conditions of the HII regions with respect to  Ly$\alpha$-emitting ones, or a larger f$_{esc}$. The latter case is consistent with a combined evolution of f$_{esc}$ and the neutral hydrogen fraction as an explanation of the lack of bright Ly$\alpha$ emission at z$>$6. A lower limit on the specific star formation rate, SSFR$>$9.1$Gyr^{-1}$ for $M_{star}=2 \times 10^9 M_{\odot}$ galaxies at these redshifts can be derived from the spectroscopically confirmed sample.
\end{abstract}


\section{Introduction}

The synergy between deep photometric and spectroscopic observations is becoming fundamental to understand the reionization epoch. On the one hand,  selection through photometric redshifts or the Lyman-break technique has enabled the determination of the evolution of the UV luminosity density and the identification of faint star-forming galaxies as the most likely responsibles of reionization  \citep[e.g.][]{Castellano2010b,Yan2011,Yan2012,Bouwens2015b,Finkelstein2015,Robertson2015,Castellano2016}. On the other hand, the spectroscopic follow-up of such photometrically selected samples has yielded constraints on the timeline of the reionization process \citep[e.g.][]{Fontana2010,Pentericci2011,Schenker2012,Caruana2012,Ono2012}.
Eventually, a thorough understanding of this major transition will require firm constraints of the physical properties of z$>$6 galaxies that affect both the interpretation of the UV LF \citep[e.g.][]{Khaire2015,Stanway2016,Wilkins2016} and the decrease of bright Ly$\alpha$ emission \citep[e.g.][]{Dijkstra2014}.
Looking for line emission signatures in broadband photometry has recently emerged as a valuable tool for investigating the evolution of galaxy properties at high-redshift \citep[][]{Smit2015b,Faisst2016}. The spectral energy distribution of objects in the reionization epoch is affected by emission from [OIII]$\lambda\lambda4959,5007$ and H$\beta$ at IR wavelengths, resulting in a bluing of the IRAC 3.6$\mu$m-4.5$\mu$m color at z$\sim$6.6-6.9, where the lines affect the 3.6$\mu$m band, and a reddening at z$>$7 when they enter the 4.5$\mu$m one \citep{Wilkins2013b}. These signatures yielded evidence of extremely strong line emission in high-$z$ galaxies, and enabled more accurate photometric redshifts and constraints on their specific star-formation rate (SSFR) \citep{Finkelstein2013,Smit2014,Smit2015a,Zitrin2015,Roberts-Borsani2015}.

In the present work we exploit deep IRAC observations to constrain the optical line emission properties of a sample of $z$-dropout galaxies from the CANDELS GOODS-South and UDS fields \citep{Grogin2011,Koekemoer2011} observed by deep spectroscopic programs aimed at detecting their Ly$\alpha$ line \citep[][and references therein, P14 hereafter, and Pentericci L. et al. 2017, in preparation]{Pentericci2014}. In Sect.~\ref{dataset} we present the sample under consideration and the procedure used to construct average (stacked) images for subsamples with different Ly$\alpha$ emission properties. The analysis of the IRAC colors in terms of optical line contribution to the broad-band photometry is given in  Sect.~\ref{lineevidence}. We discuss in Sect.~\ref{discussion} the resulting constraints on the physical properties of our targets. We present a summary in Sect.~\ref{summary}.

Throughout the paper, observed and rest--frame magnitudes are in the AB system, and we adopt the $\Lambda$-CDM concordance model ($H_0=70km/s/Mpc$, $\Omega_M=0.3$, and $\Omega_{\Lambda}=0.7$).

\begin{figure}[!ht]
   \centering{
 \includegraphics[width=7cm]{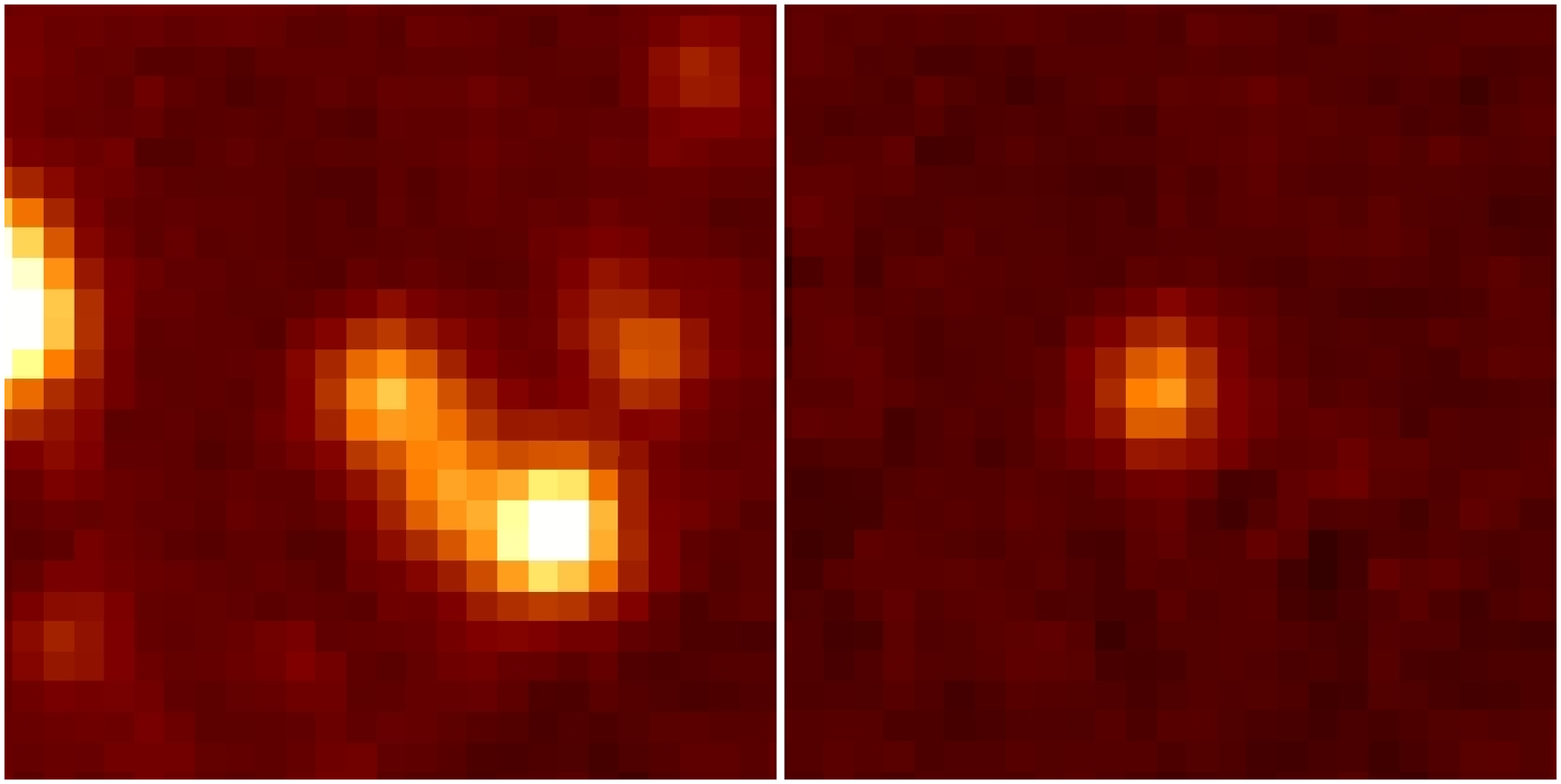}
\\
 \includegraphics[width=7cm]{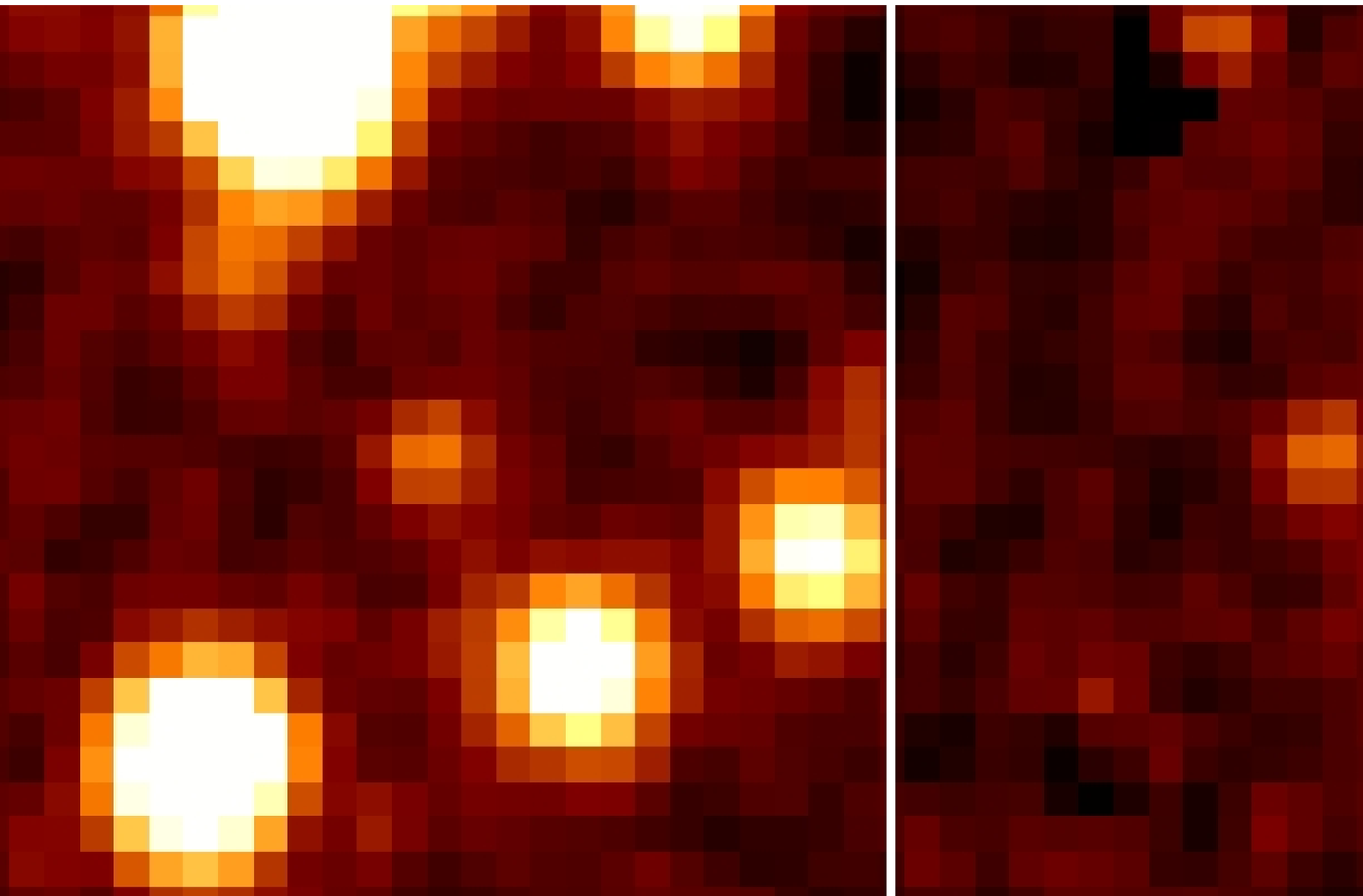}
}
   \caption{Examples from the GOODS-South (top) and UDS (bottom) fields: 15$\times$15 arcsec regions from the original CH1 map (left) compared to the residual image where all the sources surrounding the $z$-drop target have been subtracted with T-PHOT (right).}\label{fig1}%
\end{figure}
\begin{figure}[!ht]
   \centering{
 \includegraphics[width=7cm]{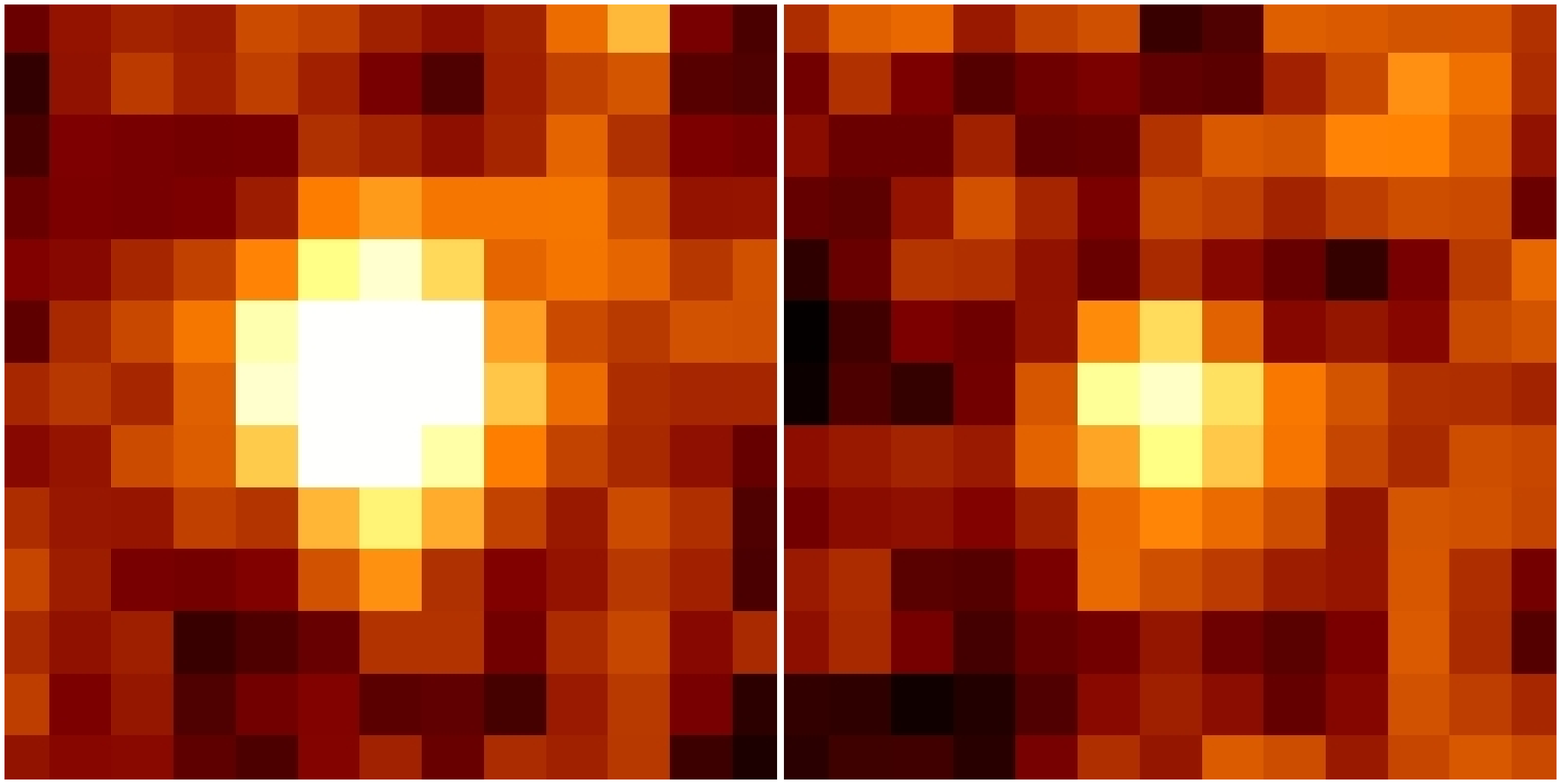}
\\
   \includegraphics[width=7cm]{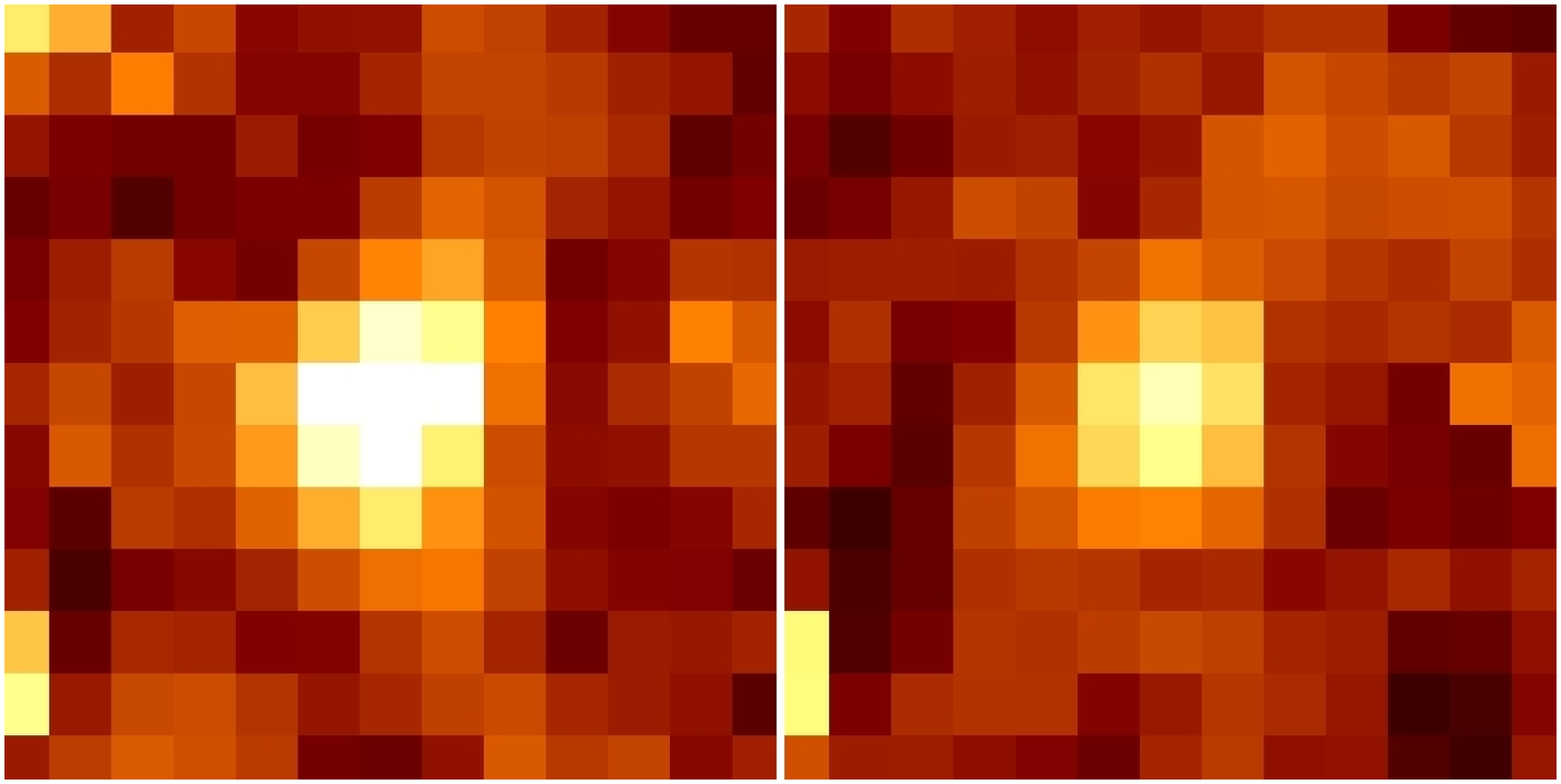}
}
   \caption{Stacked CH1 (left) and CH2 (right) thumbnails of  objects with Ly$\alpha$ emission (top) and of objects with no Ly$\alpha$ emission and 6.4$<z_{phot}<$7.0 (bottom)}\label{fig2}%
\end{figure}
\section{The High-redshift Sample}\label{dataset}

A comprehensive description of the sample will be presented in a forthcoming paper (Pentericci L. 2017, in preparation): here we summarize the information that is most relevant for the present analysis. The spectroscopic targets have been selected from the official H-band detected CANDELS catalogs of the GOODS-South \citep{Guo2013} and UDS \citep{Galametz2013} fields. Sources have been selected initially through appropriate recastings of the ``Lyman-break'' technique as described in \citet{Grazian2012}. The final color-color selection criteria take into account the different sets of passbands available in the two fields \citep[see][]{Grogin2011,Koekemoer2011} resulting in slightly different redshift selection functions \citep[Figure 1 of][]{Grazian2012}. In addition to the Lyman-break-selected candidates, we also inserted in the available FORS2 slits targets that did not pass the above criteria but had a photometric redshift of $z_{phot} > 6.5$. The photometric redshifts used for selection are the official CANDELS ones built from a set of different photo-$z$ runs through the hierarchical bayesian approach described in \citet{Dahlen2013}.  

\begin{figure*}[!ht]
   \centering{
   \includegraphics[width=7.8 cm]{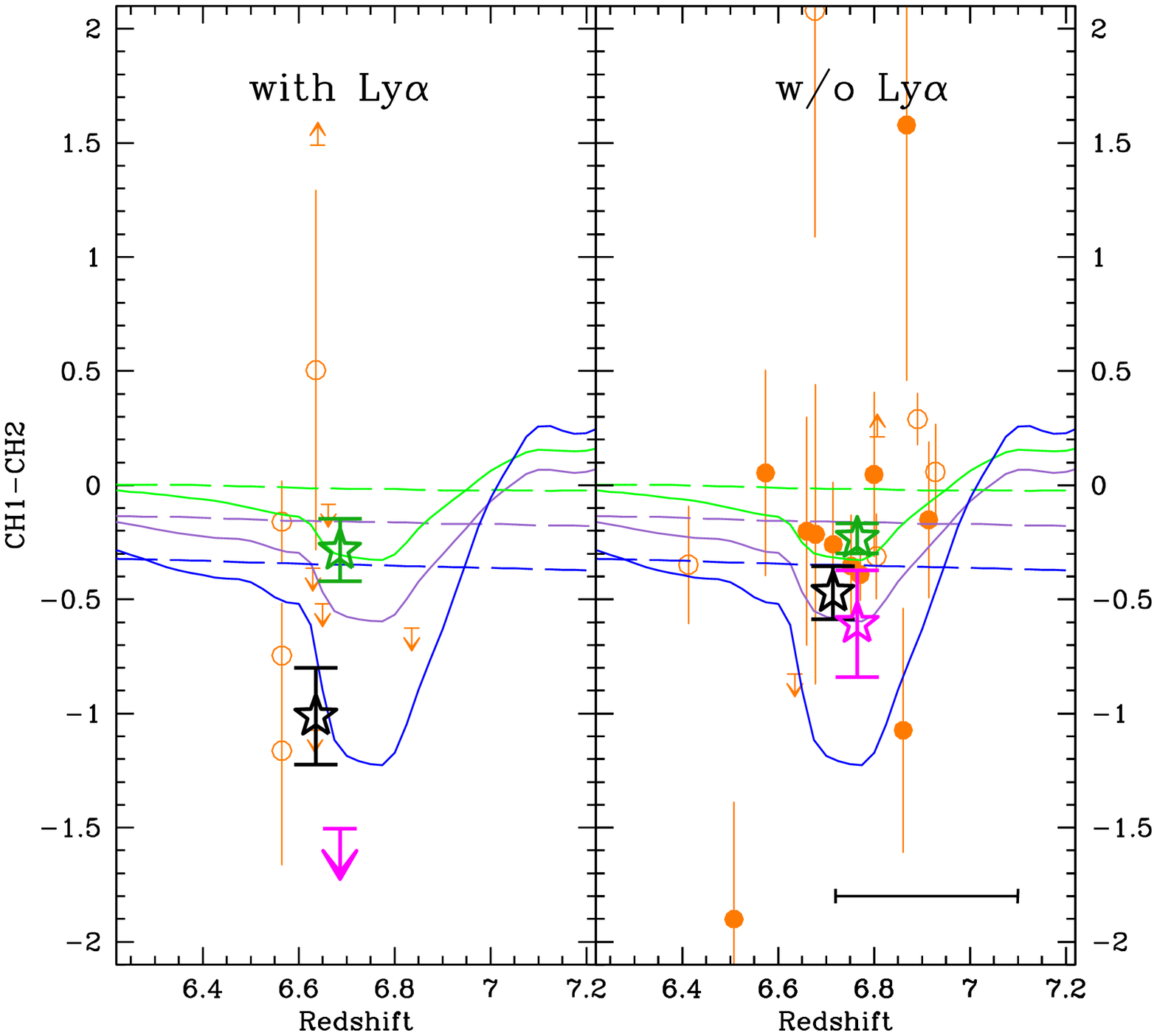}
    \includegraphics[width=7.6 cm]{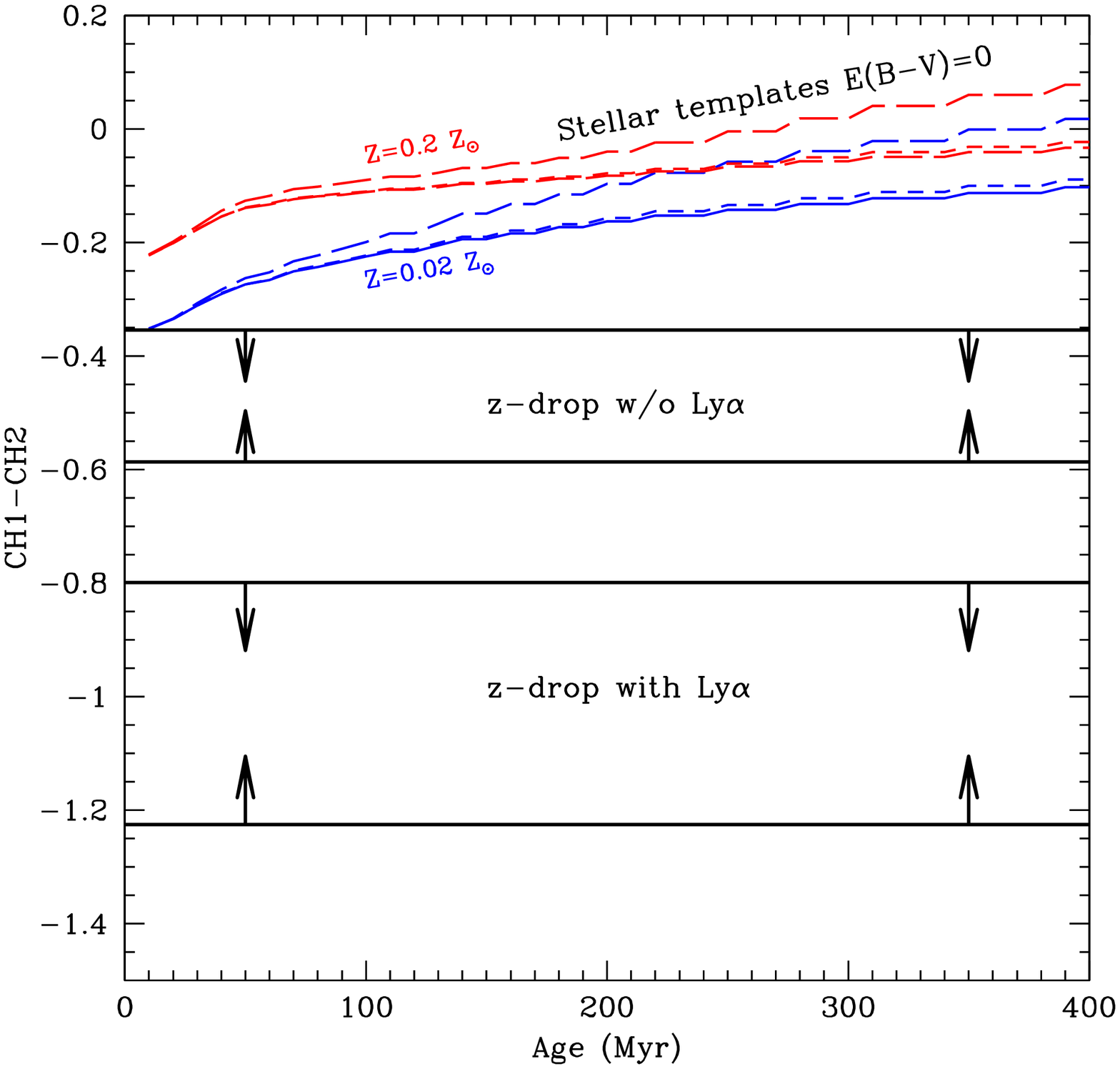}}
  \caption{\textbf{Left panel:} stacked colors (black star and errorbars) of objects with Ly$\alpha$ emission (left) and objects with no Ly$\alpha$ emission and 6.4$<z_{phot}<$7.0 (right), at the average redshift of the relevant sample. The colors of individual sources in the GOODS (UDS) field are indicated as orange filled (open) circles and errorbars. Magenta and green stars and errorbars in both panels indicate colors from stacking of the faint ($26.0<H_{160}<27.5$) and bright ($H_{160}<26.0$) subsamples, respectively (their redshift positions are slightly shifted for a better visualization). Upper and lower limits are shown as arrows. Objects undetected in both filters are not shown. The black horizontal line show the average 1$\sigma$ photo-$z$ uncertainty. The colors as a function of redshift of BC03 models including nebular emission are shown as continuous lines for objects with Age=10Myr, E(B-V)=0.0 and Z=0.02Z$_{\odot}$ (blue), Age=100Myr, E(B-V)=0.06 and Z=0.02Z$_{\odot}$ (purple),  Age=100Myr, E(B-V)=0.1 and solar metallicity (green). The case without nebular contribution is displayed by dashed lines. \textbf{Right panel:} CH1-CH2 color as a function of Age of dust-free templates at z=6.8 with metallicity Z=0.2Z$_{\odot}$ (red) and Z=0.02Z$_{\odot}$ with no nebular contribution. Continuous lines indicate models with constant SFH, exponentially declining models with $\tau$=1 and 0.1 Gyr are shown as short- and long-dashed lines respectively. Black horizontal lines bracket the 1$\sigma$ color range measured on the stacking of targets with or without Ly$\alpha$ emission as indicated by labels.}\label{fig3}%
\end{figure*}

We complemented the large program sample with data obtained by our previous programs \citep{Fontana2010,Pentericci2011}. All objects have been observed with the FORS2 spectrograph using the 600Z holographic grating (sensitivity in the range of 8000-10.000\AA~with a spectral
resolution of $R=1390$) following the observing strategy presented in P14. Finally, we add to our own sample the z$\sim$7 targets observed by ESO programmes 086.A-0968(A) and 088.A-1013(A) (P.I. Bunker) with the same FORS2 setup. The data have been processed through our own reduction pipeline, which is fine-tuned for the detection of faint emission lines \citep{Vanzella2011}.

The final spectroscopic sample comprises 84 objects including those selected only from photometric redshifts. Only 17 of them show Ly$\alpha$, in some cases quite faint, consistently with the decline of the Ly$\alpha$ emission fraction at high-redshift (Pentericci L. et al. 2017, in preparation). In the present work, we will focus on the sources in the redshift range where [OIII]+H$\beta$ generate a sharp bluing of the 3.6$\mu$m-4.5$\mu$m color. We consider 11 sources with detected Ly$\alpha$, regardless of the relevant EW, at redshift z=6.565 - 6.836 and 25 sources with no Ly$\alpha$ emission having primary photometric-redshift solution in a slightly larger range (6.4$<z_{phot}<$7.0) to conservatively account for the effect of photo-z uncertainty. The samples include galaxies with H160 spanning the range of $\sim$25.0-28.0.
We analyze the photometric properties of the spectroscopic samples exploiting the available CANDELS mosaics \citep{Koekemoer2011} in the four HST bands V606, I814, J125, and H160 that are available for both fields and the Spitzer IRAC observations in the 3.6 (CH1 herafter), 4.5 (CH2), 5.8, and 8.0 $\mu$m channels.
The IRAC mosaics of the UDS field combine observations from the SWIRE \citep{Lonsdale2003}, spUDS \citep[PI J. Dunlop,][]{Caputi2011} and SEDS  \citep{Ashby2013} surveys as described in \citet{Galametz2013}. For the GOODS-S field, we used 5.8 and 8.0 $\mu$m observations from the GOODS Spitzer Legacy project (PI: M.
Dickinson) \citep{Guo2013} together with our own reduction of all the available CH1 and CH2 IRAC observations including data from the S-CANDELS program (PI G. Fazio) \citep[see][for details]{Labbe2015,Ashby2015}. 

The analysis of individual sources is based on the 19-band photometric information for the GOODS-S and UDS fields described in \citet{Guo2013} and \citet{Galametz2013} respectively, with the notable exception of the IRAC GOODS-South photometry that we re-estimated using the full-depth maps described above\footnote{The IRAC photometry will be publicly released as part of the revised GOODS-S photometric catalog by the ASTRODEEP collaboration (Fontana et al. 2017, in preparation).}. The IRAC GOODS-S CH1 and CH2 photometry has been obtained with v2.0 of \verb|T-PHOT| \citep{Merlin2015,Merlin2016b} that exploits information from high-resolution HST images to extract photometry from lower resolution data where blending is a concern. As reference high-resolution templates, we use the sources cutouts obtained from the H160 band after dilating its segmentation map as described in \citet{Galametz2013} to recover an unbiased estimate of the total flux in the low-resolution frames. The \verb|T-PHOT| runs are performed by simultaneously fitting all of the objects in the field using object-dependent PSFs. This procedure takes into account the large variation of the point spread function resulting from the difference in position angle among the several programs contributing to the final maps. 
\begin{table}
\centering
\caption{The sample}
\label{tabsample}
\begin{tabular}{lcc}
\hline
\multicolumn{3}{c}{Ly$\alpha$-emitting sources} \\
\hline
ID$^a$ & CH1 & CH2 \\
\hline
GS\_13184 & 27.22$\pm$0.91 & $>$27.30  \\
GS\_15951 & 26.28$\pm$0.32 & $>$27.35   \\
GS\_31891 & 26.62$\pm$0.79 & $>$26.99  \\
GS\_34271 & 26.62$\pm$0.69 & $>$27.14  \\

UDS\_1920 & 24.29$\pm$0.08 & 25.04$\pm$0.21\\
UDS\_4812 & 24.55$\pm$0.13 & 25.71$\pm$0.48\\
UDS\_4872 & 24.17$\pm$0.10 & 24.33$\pm$0.15\\
UDS\_16291 & $>$27.14& 25.65$\pm$0.34\\
UDS\_19841 & 26.23$\pm$0.50 & $>$26.86\\
UDS\_23802 & 26.28$\pm$0.62 & 25.78$\pm$0.48\\
\hline
\hline
\multicolumn{3}{c}{Sources w/o Ly$\alpha$ } \\
\hline
ID$^a$ & CH1 & CH2 \\
\hline
GS\_9771 & 25.69$\pm$0.15 & 25.84$\pm$0.30\\
GS\_10377 & 24.20$\pm$0.06 & 24.60$\pm$0.10\\
GS\_13221 & $>$27.61 & $>$27.56\\
GS\_14756 & 24.99$\pm$0.20 & 26.06$\pm$0.49\\
GS\_14776 & 25.12$\pm$0.13 & 25.48$\pm$0.18\\
GS\_19483 & 25.95$\pm$0.26 & 25.91$\pm$0.25\\
GS\_20439 & $>$27.55& 27.34$\pm$0.85\\
GS\_21921 & $>$27.48& $>$27.56\\
GS\_22683 & 24.84$\pm$0.17 & 25.10$\pm$0.21\\
GS\_23182 & $>$27.45 & $>$27.53\\
GS\_24805 & 26.83$\pm$1.07 & 25.25$\pm$0.32\\
GS\_26624 & 24.34$\pm$0.08 & 26.24$\pm$0.51\\
GS\_32103 & 26.02$\pm$0.39 & 26.23$\pm$0.53\\
GS\_32516 & 26.20$\pm$0.33 & 26.40$\pm$0.38\\
GS\_33588 & $>$27.50 & $>$27.58\\
GS\_34523 & 25.58$\pm$0.34 & 25.53$\pm$0.29\\
GS\_34619 & 25.91$\pm$0.42 & $>$26.73\\

UDS\_4270 & $>$27.05& $>$26.80\\
UDS\_11752 & 24.29$\pm$0.09 & 24.61$\pm$0.16\\
UDS\_14715 & 24.74$\pm$0.12 & 25.09$\pm$0.23\\
UDS\_18014 & 24.83$\pm$0.13 & 24.77$\pm$0.16\\
UDS\_20139 & 24.33$\pm$0.08 & 24.03$\pm$0.08\\
UDS\_22859 & 26.81$\pm$0.97 & 24.73$\pm$0.19\\

\hline
\end{tabular}
\\
\smallskip
\begin{tabular}{p{5cm}}
a - progressive numbers from \citet{Guo2013} and 
    \citet{Galametz2013} for GOODS-South (GS) 
    and UDS respectively
\end{tabular}
\end{table}

\subsection{Stacking Procedure}\label{stack}
The sources under investigation have typical mid-IR flux close to the detection limit of the deep Spitzer observations. The CH1-CH2 color of the objects in our sample is in the best cases determined with an uncertainty of 0.3-0.5 magnitudes, while more than one-third of our sources have S/N$<$1 in one (mostly CH2) or both the IRAC bands. For this reason, we will base our investigation on stacked images.
We separately analyze objects with detected Ly$\alpha$ line and of those with no line detection to discern possible correlation between the optical and the Ly$\alpha$ line emission properties. We also consider subsamples of bright and faint sources to assess a possible relation between line emission properties and UV luminosity. We consider as bright objects those with $H_{160}<26.0$ (roughly corresponding to $L\gtrsim L^*$): 4 (6) sources are brighter than this limit in the Ly$\alpha$-detected (-undetected) samples respectively.
We build stacked images in the four IRAC channels and in the V606, I814, J125 and H160 HST bands. In this way, we can study the IRAC CH1-CH2 color as a probe of line emission as well as the overall ``average'' SED of the samples under consideration.
For the IRAC bands, where source confusion and blending is significant, we first perform a \verb|T-PHOT| second pass run using the option \verb|exclfile| \citep{Merlin2016b} to generate residual images where only the z$\sim$7 sources under analysis are left. In this way, all sources are modeled and those close to the z$\sim$7 are effectively removed (see, e.g. Fig.~\ref{fig1}) such that these cleaned images can be used to generate reliable stacked images of the candidates.  We then visually inspect all our sources and exclude three objects (one Ly$\alpha$ emitter and two non-emitters) due to the presence of bad residual features close to the targets that can possibly affect the photometry. In Table \ref{tabsample} we list the sources actually used for the present analysis. The stacked images are then generated as weighted averages of the individual thumbnails and are presented in Fig.~\ref{fig2}. Together with the stacks, we generate average CH1 and CH2 PSFs from the PSFs of the indivudal sources. The HST stacks are generated as weighted average images of the individual thumbnails after masking all close-by sources according to the relevant \verb|SExtractor| segmentation map. The HST photometry is obtained with \verb|SExtractor| by performing detection and estimating total magnitude in the stacked H160 band. Total magnitudes in the other bands are computed on the basis of the relevant isophotal colors with respect to the H160 one. Photometry of stacked IRAC images is estimated with \verb|T-PHOT| using the source cutout from the stacked H160 band as prior. The resulting spectral energy distributions are shown in Fig.~\ref{fig4} and Fig.~\ref{fig5} (for ``bright'' and ``faint'' subsamples).

\section{Evidence of Optical Line Emission}\label{lineevidence}

We show in Fig.~\ref{fig3} the CH1-CH2 color of Ly$\alpha$ emitting and non-emitting stacks and the colors of all individual sources under consideration. We find CH1-CH2=-1.0 $\pm$0.21 and CH1-CH2=-0.47$\pm$0.11 for Ly$\alpha$ emitting and non-emitting average sources respectively. Clearly, these colors represent the average properties of the sample.

We find that both samples show an evident relation between the UV luminosity and the CH1-CH2 color. The bright sample's stacks have a similar CH1-CH2$\simeq$-0.25 for both Ly$\alpha$ emitting and non-emitting sources.
The IRAC colors of the faint subsamples are bluer. The stacks of the faint non-emitting subsample has CH1-CH2=-0.60$\pm$0.23 while the stack of faint Ly$\alpha$-emitting sources is extremely blue (CH1-CH2$<$-1.5 at 1$\sigma$) due to the non-detection in CH2. The difference between IRAC colors of bright and faint Ly$\alpha$ emitting galaxies is signficant at the $\sim$2.5$\sigma$ level. 

As shown in Fig.~\ref{fig3}, the average negative CH1-CH2 color we find for the two samples can only be explained by the presence of optical line emission affecting the CH1 filter. The most extreme color obtained for purely stellar emission is approximately -0.35 (which would also require no dust and extreme galaxy properties, see Sect.~\ref{discussion}), much redder than the stacked color of Ly$\alpha$ emitting galaxies and only marginally compatible with the color from the stacking of non-emitting galaxies, implying that the bulk of objects in the two samples has optical line emission affecting the IRAC bands. In particular, the stacked color of the bright subsamples still suggests the presence of emission lines but is also compatible with purely stellar emission from low metallicity/low extinction galaxies, while line emission is surely present in most of the objects contributing to the faint subsamples. Interestingly, this value is bluer than for the youngest and lowest metallicity templates in our library suggesting that the physical conditions in distant HII regions can be more extreme than what is assumed in our nebular emission model \citep{Schaerer2009}.

\begin{figure}[!ht]
   \centering{
   \includegraphics[width=6.5 cm, angle=-90]{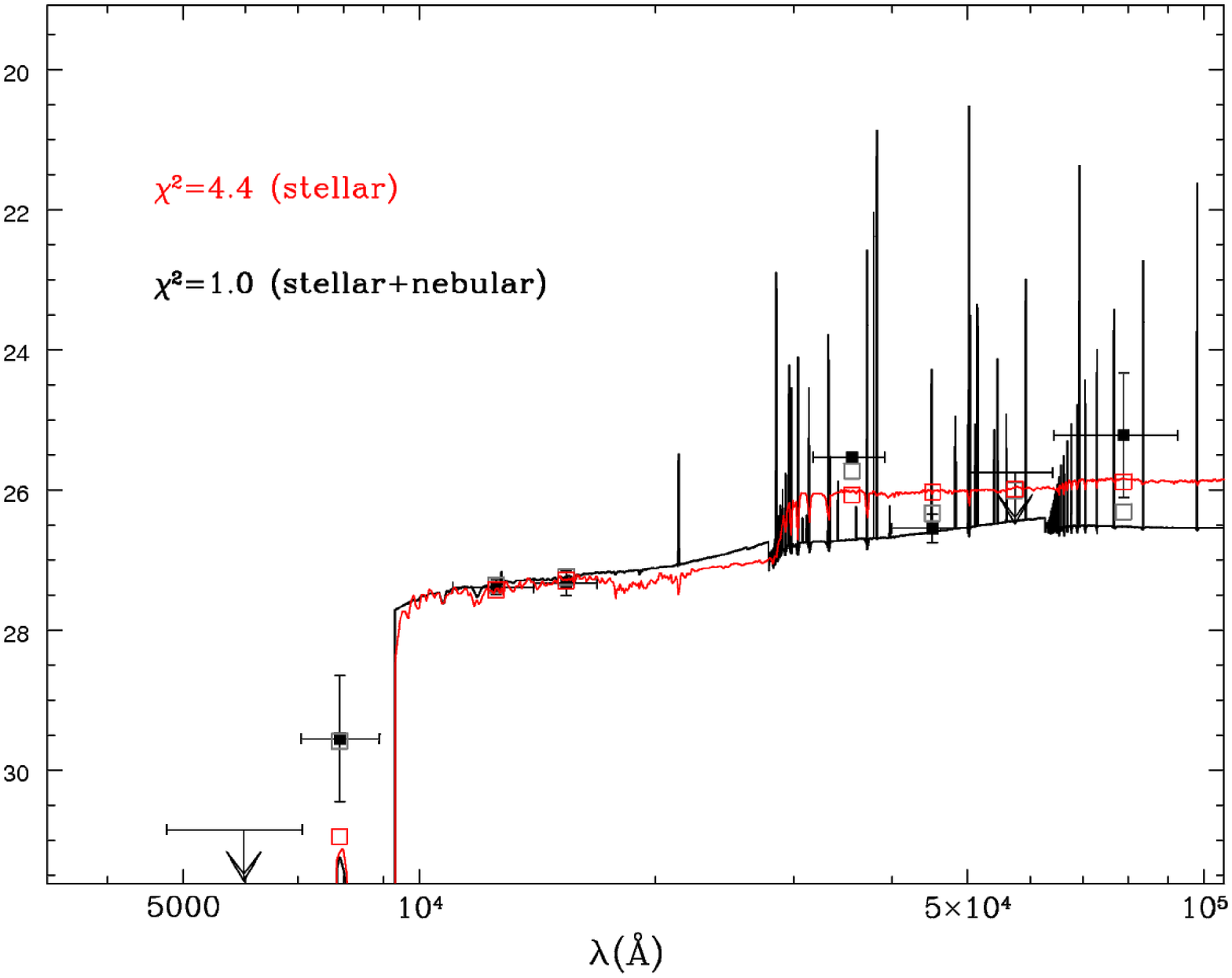}
   \includegraphics[width=6.5 cm, angle=-90]{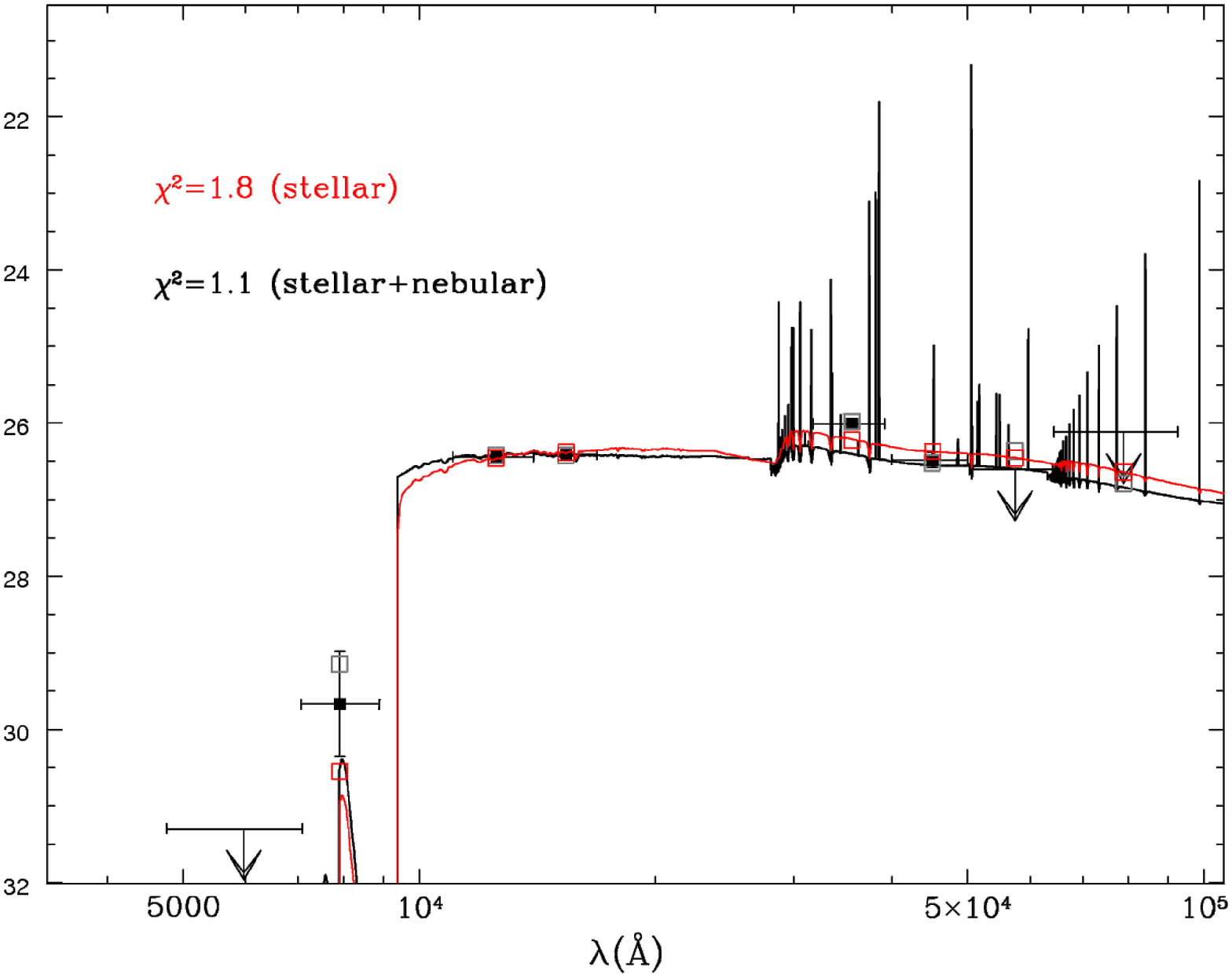}
}
  \caption{Spectral energy distribution of the stacking from sources with Ly$\alpha$ emission (top) and objects with no Ly$\alpha$ emission (bottom). Best-fit SEDs from stellar-only and stellar+nebular fits are shown in red and black, respectively, with relevant best-fit magnitudes in the different bands shown as empty squares. Filled squares and errorbars indicate the measured photometry.}\label{fig4}%
\end{figure}

\begin{figure*}[!ht]
   \centering{
   \includegraphics[width=12.5 cm]{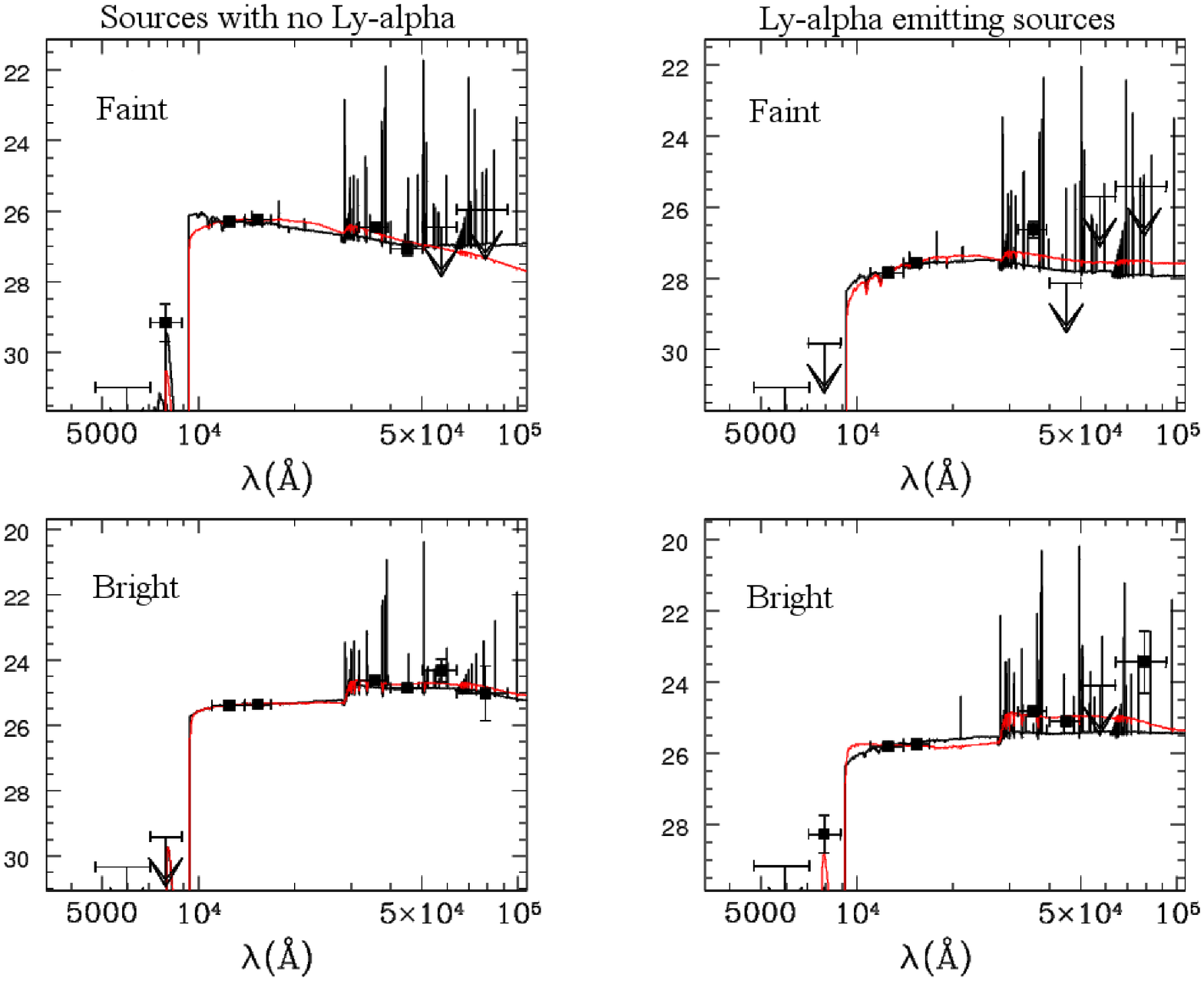}
}
  \caption{Spectral energy distribution of the stacking from sources with faint (top panels) and bright (lower panels) sources in the Ly$\alpha$ emitting (left) and non-emitting (right) subsamples.}\label{fig5}%
\end{figure*}

The evidence of optical emission lines is also shown by a SED-fitting of the stacked multi-band photometry. We fit the eight-band photometry with our $\chi^2$ minimization code \citep{Fontana2000} fixing the redshift at the average one of the relevant sample. The fit is performed both with stellar only templates from the library of \citet{Bruzual2003} (BC03 hereafter)  and also including the contribution of line emission as in \citet{Schaerer2009} assuming an escape fraction of ionizing photons f$_{esc}$=0 \citep[see also][]{Castellano2014}. A comparison of the stellar and stellar+nebular fits shows that the former solution is disfavored in terms of $\chi^2$ (Fig.~\ref{fig4}) By varying the contribution of nebular emission from f$_{esc}$=0 to f$_{esc}$=1 at 0.2 steps, we find that f$_{esc}$=0 models are always favored. Templates with f$_{esc}>0.4$ are excluded at 1$\sigma$ in the case of Ly$\alpha$-emitting galaxies, while the difference in terms of $\chi^2$ among the various templates is not significant in the case of Ly$\alpha$-undetected ones.
Considering that photometric redshift estimates do not rely on nebular templates, the evident nebular feature in the IRAC bands of sources with no Ly$\alpha$ redshift, together with the deep non-detection in the stacked V606 band ($>$31.4 mag at 1$\sigma$), provides further evidence that these objects are robust z-dropout galaxies thus strengthening the case for a declining Ly$\alpha$ fraction at z$>$6.

\section{Discussion}\label{discussion}
We can convert the observed IRAC color into a combined rest-frame EW([OIII]+H$\beta$) by assuming a baseline color for the intrinsic stellar emission, which, in turn, depends on age, E(B-V) and metallicity of the stellar population. Intrinsic colors range from CH1-CH2$\simeq$-0.35 for a dust-free Z=0.02$Z_{\odot}$ template of Age=10Myr, to CH1-CH2$\gtrsim$0.2 (e.g. Age=100Myr, E(B-V)=0.2, solar metallicity). 
In particular, age and dust extinction are the factors that mostly affect the continuum shape, with a 0.2 mag color difference between templates at E(B-V)=0 and E(B-V)=0.15 (at fixed age and metallicity) and between templates at Age=0 and Age=300 Myr (at fixed dust extinction and metallicity). A 0.1 mag difference in color is found between Z=0.02$Z_{\odot}$ and solar metallicity templates of similar age and dust extinction. In principle, the difference between the IRAC colors of Ly$\alpha$-emitting and non-emitting galaxies ($\sim$0.5 mag) can be completely explained by a difference in the underlying stellar optical continuum with Ly$\alpha$-emitting being very young, metal-poor, and dust-free, and objects lacking Ly$\alpha$ emission being $>$100Myrs old, metal enriched, and mildly extincted. However, the typical UV slopes obtained from the J125-H160 stacked photometry is $\beta\simeq -1.9$ for both samples and the distribution of individual UV slope in the two samples is similar (Pentericci L. et al. 2017, in preparation). We can thus exclude a significant presence of dust-free low metallicity galaxies among Ly$\alpha$-emitting galaxies since such an extreme population would show a bluer slope $\beta\sim$-2.7 \citep[considering also contribution from nebular continuum, see, e.g.][]{Castellano2014}. Therefore, different physical properties can contribute, but not completely explain, the difference between IRAC colors in our samples. For simplicity, we consider a flat CH1-CH2=0.0 for the no-emission-line case, as expected for a reference 100Myr old Z=0.2$Z_{\odot}$ galaxy with UV slope $\beta\sim-1.9$ (corresponding to E(B-V)$\sim$0.12), to convert IRAC colors into equivalent widths of the optical line emission. The measured color term can be then converted into EW([OIII]+H$\beta$) = 1500$^{+530}_{-440}$ \AA~(Ly$\alpha$ emitting sample) and EW([OIII]+H$\beta$)=520$^{+170}_{-150}$ (Ly$\alpha$ undetected sample). These values are consistent within the uncertainty with the line strength predicted by the stellar+nebular SED-fitting on the eight-band stacked photometry, thus providing further evidence that a difference in the stellar SEDs is unlikely to explain the different IRAC colors. The stacking of bright sources yield EW([OIII]+H$\beta$)$\sim$230-290\AA. The largest equivalent widths are obtained for the faint subsamples with EW([OIII]+H$\beta$)=720 $^{+400}_{-330}$ \AA~of Ly$\alpha$-undetected sources and a lower limit of EW([OIII]+H$\beta$) $>$2900\AA~ of Ly$\alpha$ emitting ones. In fact, given the similar color of the stacked bright subsamples, the difference between Ly$\alpha$ detected and undetected objects appears to be mostly confined to the subsamples of faint ($H160>26.0$) sources. We summarize in Table~\ref{tabstack} measurements for the different subsamples.
\begin{table*}
\centering
\caption{Stacked IRAC photometry}
\label{tabstack}
\begin{tabular}{lcc}
\hline
Subsample & CH1-CH2 &  EW([OIII]+H$\beta$) (\AA) \\
\hline
Ly$\alpha$-emitting, all& -1.0$\pm$0.21 & 1500$^{+530}_{-440}$ \\
Ly$\alpha$-emitting, bright& -0.28$\pm$0.14 & 290$^{+170}_{-150}$ \\
Ly$\alpha$-emitting, faint& $<$-1.5 & $>$2900 \\
No Ly$\alpha$, all& -0.47$\pm$0.11 & 520$^{+170}_{-150}$ \\
No Ly$\alpha$, bright&  -0.23$\pm$0.07 &  230$^{+70}_{-70}$ \\
No Ly$\alpha$, faint& -0.61$\pm$0.23 & 720$^{+400}_{-330}$ \\
\hline
\end{tabular}
\end{table*}
Notably, the different IRAC colors of bright and faint Ly$\alpha$ emitting galaxies ($\sim$1 mag) cannot be explained from a variation of the underlying stellar continumm alone ($\lesssim$0.5 mags). The relation between EW([OIII]+H$\beta$) and UV luminosity, which is evident in both Ly$\alpha$-detected and -undetected samples can also be explained by a relation between age and UV luminosity. Moreover, such bright optical line emission from sub-L$^*$ sources implies that stellar feedback is either not strong enough to deplete their inter-stellar medium or the sources are too young, and thus feedback has not been effective for a long enough time to affect the ISM.

An intriguing possibility is that different physical properties of the HII regions concur in explaining both Ly$\alpha$ visibility and a larger EW([OIII]+H$\beta$) \citep[see also][]{Roberts-Borsani2015,Stark2016}. The IRAC color of the Ly$\alpha$ emitting galaxies can thus be explained by these objects being younger and more metal poor, and thus with harder ionization fields, than non-emitting ones. A higher escape fraction of ionizing photons can also explain a lower EW of the optical emission lines and play a role in the low Ly$\alpha$ visibility. In the next section, we will discuss the relation between physical conditions of the HII regions and EW([OIII]+H$\beta$). 

An alternative explanation of the difference between the two stacks can be uncertainties affecting the sample of objects with no Ly$\alpha$. We can exclude with high confidence any contamination from low-redshift interlopers since no other lines are detected in any of the objects (P14, Pentericci L. et al. 2017, in preparation) and also because of the mag$>$31.4 non-detection on the stacked V606 band. Moreover, the nebular feature typical of this redshift range is more evident for faint sources where a larger contamination might be expected given the lower reliability of photometric redshifts. However, we can not exclude the possibility that the Ly$\alpha$-undetected samples in the 6.4$<z_{phot}<$7.0 range actually contain sources with true redshift $>$7.0, which would partially erase the line signature. At z$>$7.0 the CH1-CH2 can be as red as $\sim$0.5-0.8 \citep[e.g.][]{Roberts-Borsani2015} because of [OIII]+H$\beta$ affecting the 4.5$\mu$m passband: this can be the case of some of the sources in our sample with a positive color term (Fig.~\ref{fig3}). Similarly, H$\alpha$ emission can add to the CH2 flux of objects at z$\sim$6.5. In such a case, the EW([OIII]+H$\beta$) we measure for Ly$\alpha$-undetected sources should be considered to be a lower limit of the real, typical line strength. 

We perform two tests to ascertain possible biases due to the photometric-redshift selection. First of all, we restrict the analysis to a more conservative range 6.6$<$z$<$6.9 and excluding sources with red IRAC colors (CH1-CH2$>$1): we find an average CH1-CH2$\sim$-0.2 again suggestive of low EW([OIII]+H$\beta$). As a second test, we inspected the photometric-redshift probability distribution functions of our objects to isolate those with highest probability (p$>$0.75) of being in the 6.6$<$z$<$6.9 range. Four out of five objects have IRAC color in the range of $\sim$-0.26 to -0.39 the remaining one being UDS\_22859 with CH1-CH2$\sim$2. These results suggest no obvious bias due to photometric-redshift selection in the result from the stack of Ly$\alpha$-undetected sources, though a future spectroscopic detection of optical lines themselves with JWST is likely the only way to overcome the effect of photometric-redshift uncertainties in this kind of analysis.

\begin{figure}[!h]
   \centering{
    \includegraphics[width=6.5 cm, angle=-90]{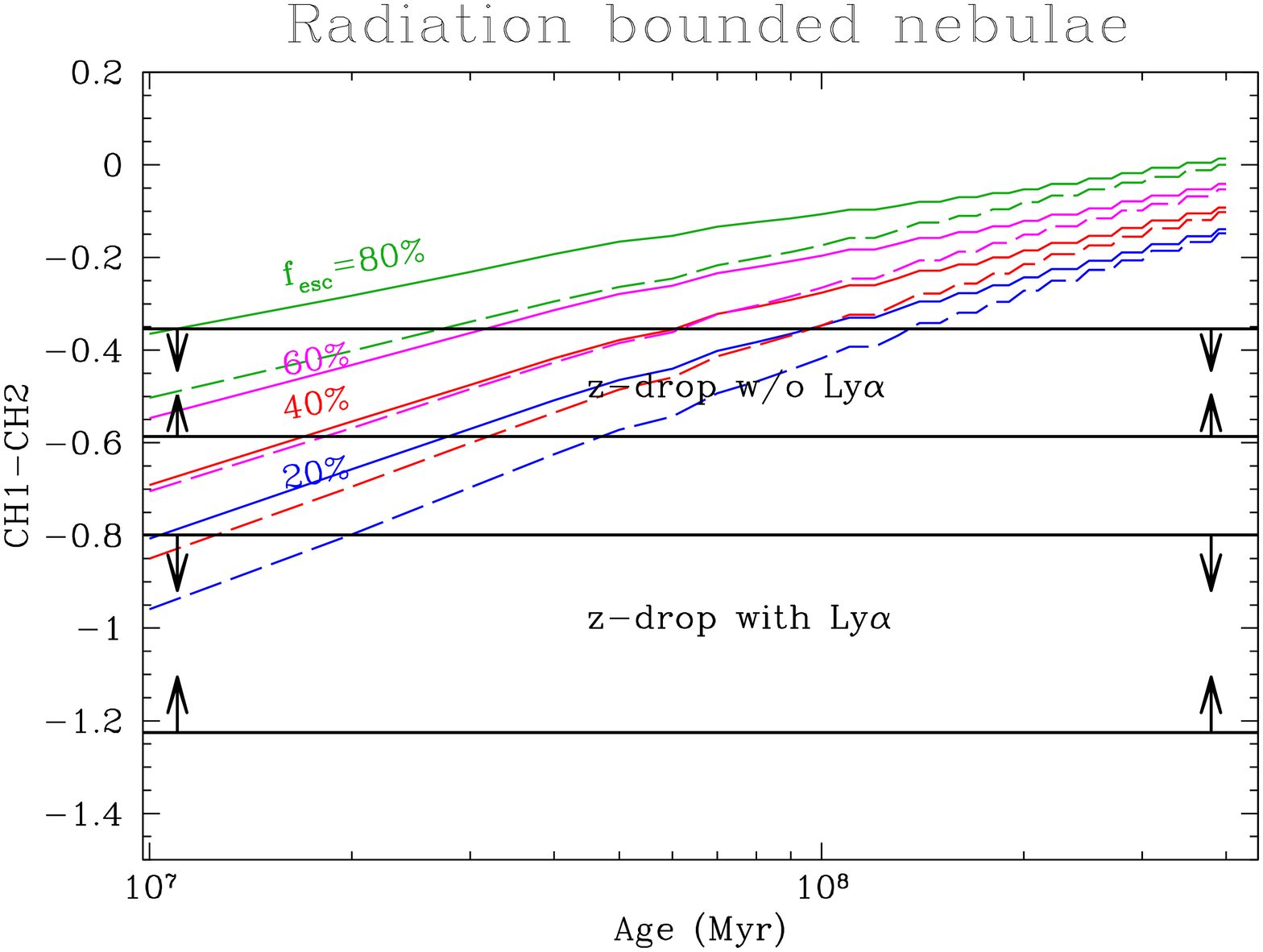}
    \includegraphics[width=6.5 cm, angle=-90]{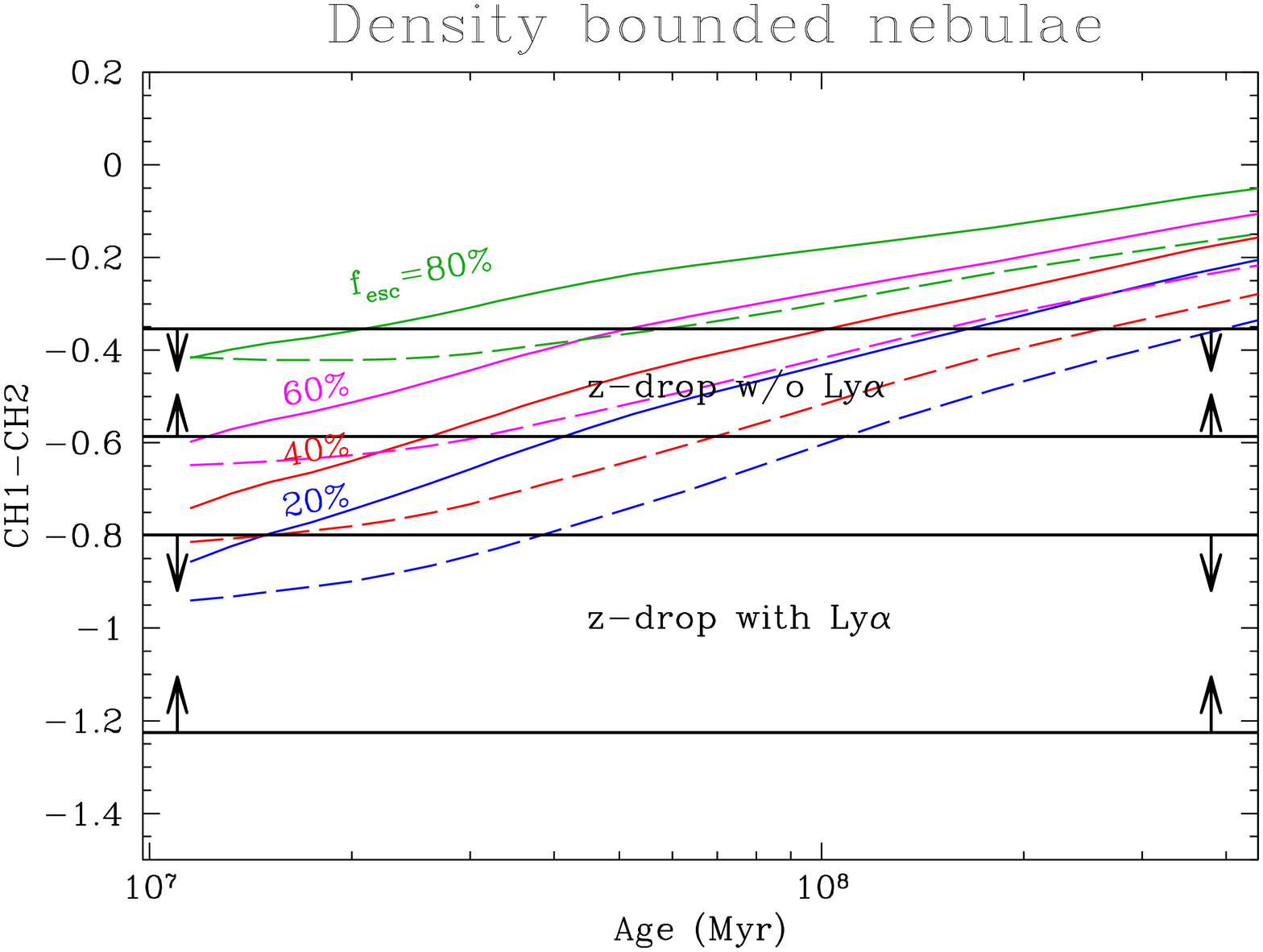}}
  \caption{Same as the right panel of Fig.~\ref{fig2} but for models with different escape fraction of ionizing photons (see labels) assuming emission exclusively comes from either radiation bounded (top) or density bounded nebulae (bottom). Continuous and dashed lines indicate models at z=6.8 with Z=0.2$Z_{\odot}$ and Z=0.02$Z_{\odot}$ respectively.}\label{fig6}%
\end{figure}

\begin{figure}[!h]
   \centering{
    \includegraphics[width=9 cm]{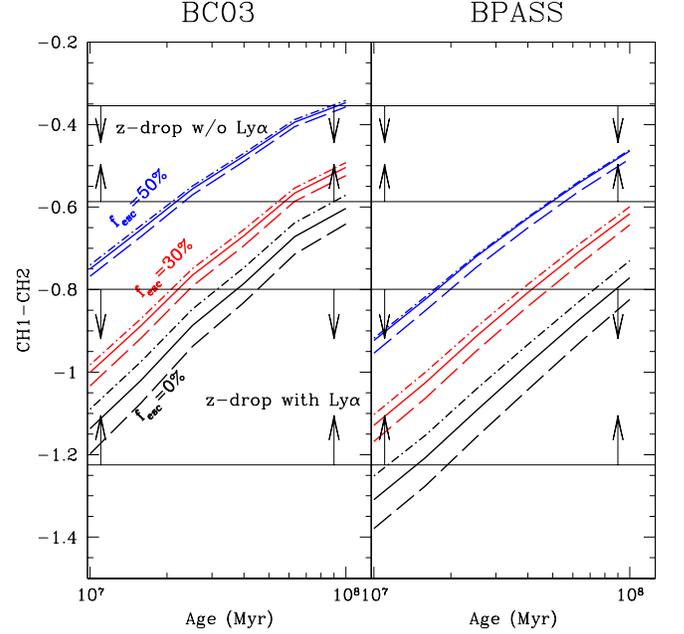}}
  \caption{IRAC color as a function of galaxy age for HII regions at different physical conditions and escape fractions using ionizing emission computed from BC03 (left panel) and BPASS (right) templates using CLOUDY. Dashed, continuous, and dotted-dashed lines indicate models with $(log(q/cm~s^{-1}),n(cm^{-3}))= (9,10^3), (9,10^2)$ and $(7.75,10^3)$ respectively. Black, red, and blue set of lines assumed f$_{esc}$=0.0,0.3 and 0.5 respectively. A metallicity of Z=0.2$Z_{\odot}$ is assumed in all cases.}\label{fig7}%
\end{figure}

\subsection{Implications on the Escape Fraction}

The escape of ionizing Lyman continuum (LyC) radiation from star-forming regions affects nebular emission and line strength.  In particular, a high escape fraction and a high neutral hydrogen fraction in the IGM have a similar effects on Ly$\alpha$ visibility \citep{Hutter2014,Hutter2015} while optical emission lines such as O[III] and H$\beta$ are only affected by f$_{esc}$. \citet{Dijkstra2014} found that the observed decline of the Ly$\alpha$ emission at high-redshift can be explained by a small increase of the Lyman continuum escape fraction $\Delta$f$_{esc}<$0.1 assuming f$_{esc}$ is already high ($\sim0.65$) at z=6, or by a modest increase in both the escape fraction ($\Delta$f$_{esc}\simeq$0.1) and the neutral IGM fraction ($\Delta \chi_{HI} \simeq 0.2$) from z=6 to z=7 starting from a f$_{esc}=0.15$ at z=6.  

Two mechanisms can be responsible of LyC leakage: the presence of ``holes'' in standard radiation-bounded HII nebulae, and the formation of incomplete Str\"{o}mgren spheres, or ``density bounded'' HII regions \citep[][Z13 herafter]{Zackrisson2013}. Real cases of LyC leakage are most probably due to a combination of the two phenomena.  As discussed in depth by Z13, a combined measurement of the UV slope and of EW(H$\beta$), which will become feasible only with JWST, yield to general constraints on the escape fraction of ionizing photons from high-redshift galaxies, albeit mid/far-IR rest-frame information might be needed to disentangle the effects of dust. However, the present evidence of strong line emission affecting the broadband colors of high-redshift galaxies allows us to put first constraints on the LyC leakage since line luminosity is suppressed at increasing f$_{esc}$ with no line emitted in the extreme case of f$_{esc}$=1.
We compute the expected IRAC color for different f$_{esc}$ values as a function of galaxy age in two different ways: (1) from stellar+nebular templates following \citet{Schaerer2009}, where hydrogen lines are computed considering case B recombination, and relative line intensities of He and metals as a function of metallicity are taken from \citet{Anders2003} and assumed to be independent of f$_{esc}$, as expected in ionization bounded nebulae; (2) by modeling a density bounded nebula with \verb|CLOUDY| \citep{Ferland1998,Ferland2013} adopting the same assumption as described in \citet{Nakajima2014} and fixing the ionization parameter at $log(q/cm~s^{-1})=7.75$.
Stellar templates from the BC03 library and a constant SFH with a minimum age=10Myr are assumed in both cases and considering E(B-V)=0.15 (for Z=0.02$Z_{\odot}$) and E(B-V)=0.10 (Z=0.2$Z_{\odot}$) because this is the lowest value allowed by the observed UV slope at age=10Myr, where line EW is the largest for any f$_{esc}$.

In Fig.~\ref{fig6} we compare the observed stacked colors of our samples with the color predicted for our reference models of radiation bounded (top panel) and ionization bounded (bottom) nebulae. In both cases, we find that the EW([OIII]+H$\beta$) of the Ly$\alpha$-emitting stack is best reproduced by models with null escape fractions: it is consistent with f$_{esc}$ up to 20\% but only for extremely young and probably unrealistic ages, especially in the radiation bounded nebulae scenario ($\sim$10Myrs). On the other hand, the CH1-CH2 color of the Ly$\alpha$ undetected stack is compatible with a larger f$_{esc}$ from very young and metal poor galaxies, or with a similar f$_{esc}<$20-40\% for ages up to $>$100 Myrs in the density bounded case. 
We further explored how different physical conditions in the HII regions can effect the emission line strenght and thus the IRAC colors, using CLOUDY. In particular, since it has been suggested that high-redshift star-forming regions might be characterized by more extreme conditions   \citep[e.g.][]{Shirazi2014,Nakajima2014,Nakajima2016}, we assume a harder ionization field $log(q/cm~s^{-1})=9.0$ and a higher density n=1000cm$^{-3}$. The results relevant to the present case are shown in the left panel of Fig.~\ref{fig7}, a thorough investigation of the ISM conditions will be presented in a forthcoming paper (De Barros S. et al. 2017 in preparation). We find that for Ly$\alpha$-emitting sources f$_{esc}>$50\% can still be excluded at any age, while they are compatible with f$_{esc}\lesssim$30\% at young ages ($<$20Myr).  We have then performed the same calculation described above using templates from the BPASSV2.0 library including the effect of interacting binary stars that can also significantly affect the emission budget of ionizing photons at high-redshift \citep{Eldridge2009,Stanway2016}. As shown in the right panel of Fig.~\ref{fig7}, the boosted ionizing flux in the BPASS templates yield $\sim$0.2 mag bluer colors than BC03 ones. In any case, even in the most favorable ionizing conditions, we can basically exclude an escape fraction larger than 50\%  at all ages. This test highlights that not only a variation in the escape fraction but also different physical properties of the HII regions can contribute in explaining the different IRAC color of Ly$\alpha$-detected and Ly$\alpha$-undetected sources.  Clearly, only future spectroscopic investigations of the optical rest-frame emission will be able to assess the physical conditions of primordial HII regions and the link between Ly$\alpha$ emission, gas properties and f$_{esc}$.
If confirmed, a larger f$_{esc}$ in Ly$\alpha$ undetected sources would provide evidence of a scenario with a milder evolution of the neutral hydrogen fraction as suggested by \citet{Dijkstra2014}. In particular, the  ``density bounded'' leakage case can be probed by future JWST mid-infrared spectroscopic observations disentangling the strong combined EW([OIII]+H$\beta$) detected in these galaxies to look for non-standard [OIII]/H$\beta$ and [OIII]/[OII] ratios as indirect tracers of high f$_{esc}$ \citep[e.g.][and references therein]{deBarros2016,Vanzella2016}. Interestingly, the similar EW([OIII]+H$\beta$) inferred for bright sources regardless of their Ly$\alpha$ emission suggests that f$_{esc}$ or physical differences might involve only sub-$L^*$ galaxies while other factors, including IGM transmission, affect the Ly$\alpha$ visibility of bright ones.

\subsection{The Specific Star-formation Rate of Reionization Galaxies}
Our sample of spectroscopically confirmed high-redshift sources allows us, for the first time, to constrain the SSFR during the reionization epoch from a homogeneously selected sample of objects with secure redshift. On the one hand, the strength of the optical line emission can be used as a star-formation rate indicator. On the other hand, the continuum emission in the 4.5$\mu$m band corresponds to the optical rest-frame emission and can be used as a proxy of the total stellar mass. As a first estimate, we compute a conservative lower limit on the SSFR, solely based on the stacked IRAC photometry \citep[e.g.][]{Smit2014}.
We first build a library of constantly star-forming models from both the BC03 and BPASSV2.0\footnote{We compute the mass normalization of BPASSV2.0 constant SFR templates assuming a 30\% mass fraction recycled in the ISM \citep[e.g.][]{Cole2000,Renzini2016}.} libraries at different ages that we use as a reference to estimate SFR and stellar mass. We assume a Salpeter IMF and consider models with E(B-V) from 0 to 1. and metallicity Z=0.02,0.2,1.0 $Z_{\odot}$ (for BC03) or Z=0.001,0.004,0.02 (BPASSV2.0). The SFR is obtained from the IRAC color after converting the corresponding EW([OIII]+H$\beta$) into H$\alpha$ luminosity assuming standard line ratios \citep{Anders2003} and a redshift of z=6.7, which is the average value of the Ly$\alpha$-detected sample. Stellar mass is obtained by computing the relevant conversion with respect to the mid-IR continuum luminosity probed by the CH2 band. Among all considered models, we look for the one yielding the lowest SSFR that we can safely assume as a conservative lower limit for the typical SSFR at these redshifts. We find minimum values of SSFR=9.1 Gyr$^{-1}$ and SSFR=10.5 Gyr$^{-1}$ from BC03 and BPASS models, respectively, with a stellar mass of $\sim$2$\times 10^9 M_{\odot}$. Our analysis points to a larger SSFR with respect to the previous estimate from \citet{Smit2014} who used emission line signatures in seven LBG candidates at $z\sim$6.6-7.0 to derive a lower limit of 4 Gyr$^{-1}$. 

An increased SSFR in low luminosity galaxies might explain the difference between \citet{Smit2014} (focused on  L$>$L$^*$ sources) and our sample that includes fainter galaxies. In turn, this can be related to the bimodality found in z$\sim$5-7 galaxies by \citet{Jiang2016} with ``old'' (age$>$100Myr) having SSFR$\sim$3-4 Gyr$^{-1}$, and young (age$<$30Myr) having ten times larger SSFR.
The real specific star-formation rate can be much higher than this limit. In fact, the nebular-stellar fit of the stacked SED yields an SSFR$=$103$^{+35}_{-39} Gyr^{-1}$  \citep[stellar mass in the range $M_{star}=0.4-0.6 \times 10^9 M_{\odot}$ assuming an initial mass function from][]{Salpeter1955}, which is consistently a factor $\sim$2 higher than the corresponding SSFR$\sim$50 $Gyr^{-1}$ found by \citet{Smit2014}, but similar to the SSFR of low-mass z$>$3 galaxies measured by \citet{Karman2016}. We note that the SSFR we find for our Ly$\alpha$-emitting z$\sim$7 sources is comparable to  estimates from other spectroscopically confirmed galaxies at z$\gtrsim$7, ranging from $\sim$10 to 20 $Gyr^{-1}$ \citep{Oesch2015,Song2016,Stark2017} to values $>$100 $Gyr^{-1}$ \citep{Finkelstein2013,Huang2016}. High SSFR at these redshift are also favored by the z$\sim$3-6 redshift trend presented in \citet{deBarros2014}.

\section{Summary and Conclusions}\label{summary}
We have analyzed the IRAC 3.6$\mu$m-4.5$\mu$m color to gather information on optical line emission of a sample of z$\sim$7 galaxies in the CANDELS GOODS and UDS fields that have been targeted by a spectroscopic campaign to detect their Ly$\alpha$ line. After dividing the sample into Ly$\alpha$-detected (10 sources) and -undetected (23 sources at 6.4$<z_{phot}<$7.0) subsamples, we built stacked images in the V606, I814, J125, and H160 HST bands and in the four IRAC channels at 3.6-8.0 $\mu$m. We analyzed the SEDs and the colors of the stacked sources finding the following.

\begin{itemize}
\item There is evidence of strong [OIII]+H$\beta$ emission in the average (stacked) SEDs both of galaxies with detected Ly$\alpha$ emission and of those lacking Ly$\alpha$ line. On the basis of the $\chi^2$, the SED-fitting including nebular contribution is clearly preferred with respect to stellar-only models. The stacked V606  band from objects lacking Ly$\alpha$ line confirms the reliability of these sources as high-redshift candidates through a deep non-detection at mag$>$31.4, corresponding to a V606-H160$\simeq$5. 
\item The CH1-CH2 color is bluer (-1.0 $\pm$0.21) for the average object with a detected Ly$\alpha$ line than for non-emitting sources (-0.47$\pm$0.11). The IRAC colors can be translated into equivalent width EW([OIII]+H$\beta$) = 1500 $^{+530}_{-440}$ \AA~ (Ly$\alpha$ emitters) and EW ([OIII]+H$\beta$) = 520 $^{+170}_{-150}$ \AA~(non-emitters) assuming a flat intrinsic stellar continuum. Optical emission lines appear stronger in the subsamples of faint ($26.0<H_{160}<27.5$) objects, with the average color of bright ($H_{160}<26.0$) sources compatible with stellar-only emission from low metallicity young galaxies. Bright galaxies with and without confirmed Ly$\alpha$ emission show similar CH1-CH2 colors, such that the difference between the two populations effectively lies in the faint subsamples.
\item The different IRAC color between the two populations can be most likely explained by a difference in physical conditions of the HII regions, with Ly$\alpha$-emitting galaxies being younger and/or more metal poor, thus with harder ionization fields, or by a larger escape fraction in non-emitting sources.  A possible dilution of the line signature due to z$>$7 galaxies in the photometric-redshift sample cannot be excluded.
\item The strong signature of optical line emission of Ly$\alpha$ detected objects yield to f$_{esc}\lesssim$20\% on the escape fraction of ionizing photons from these objects both in the case of radiation bounded and of density bounded HII regions. A larger f$_{esc}$ limit ($\lesssim$50\%) is found when assuming the extreme case of very high density and ionization parameter and the contribution from interacting binaries to the ionizing flux. The optical line emission from Ly$\alpha$ undetected sources can be explained by a larger f$_{esc}$ from very young and metal poor galaxies, or with a similar f$_{esc}<$20-40\% for ages up to $\sim$80-130Myr. These results are qualitatively in agreement with the scenario suggested by \citet{Dijkstra2014} of a combined evolution of f$_{esc}$ and neutral hydrogen fraction explaining the lack of bright Ly$\alpha$ emission at z$>$6.
\item By using only the spectroscopically confirmed objects, we derive SSFR$=$ 103$^{+35}_{-39} Gyr^{-1}$ for $M_{star}=5 \times 10^8 M_{\odot}$ galaxies at z$\sim$6.7 from the stacked SED, and a robust lower limit of SSFR$=$9-10$Gyr^{-1}$ (depending on the assumed library) under the most conservative assumptions on the conversion factor used to derive SFR and stellar mass using only information from the mid-IR photometry.
\end{itemize}

Mid-IR spectroscopy with JWST is clearly needed to move beyond constraints from broadband observations. In this respect, it is interesting to note that the strength of the optical line signature found in our sample implies typical [OIII] and H$\beta$ fluxes of $\sim 10^{-16}-10^{-17} erg/s/cm^2$. Such bright lines can be detected at high S/N by NIRspec with few minutes of integration time\footnote{https://jwst.etc.stsci.edu} allowing us to fully constrain the dependence of Ly$\alpha$ emission on physical properties and to look for unusual line ratios as a signature of large escape fraction from density bounded regions.

\acknowledgments
We thank R.J. McLure for kindly providing IRAC mosaics of GOODS-South, and J.J. Eldridge for assistance in using the BPASS library. S.D.B. thanks Gary Ferland and the organizers of the 2015
Belfast CLOUDY Winter School as well as Kimihiko Nakajima for their support regarding CLOUDY simulations. K.C. acknowledges funding from the European Research Council through the award of the Consolidator Grant ID 681627-BUILDUP. The research leading to these results has received funding from the European Union Seventh Framework Programme (FP7/2007-2013) under grant agreement n° 312725. This work is based on data obtained with ESO Programmes 084.A-0951, 085.A-0844, 086.A-0968, 088.A-1013, 088.A-0192, and 190.A-0685.

\end{document}